\documentclass{IEEEtran}
\usepackage{cite}
\usepackage{amsmath,amssymb,amsfonts}
\usepackage{algorithm}
\usepackage{algorithmic}

\usepackage{graphicx}
\usepackage{textcomp}
\usepackage{comment}
\usepackage{booktabs}
\newcommand{\ra}[1]{\renewcommand{\arraystretch}{#1}}
\usepackage{threeparttable}
\usepackage{xcolor}
\usepackage{soul}

\renewcommand{\st}[1]{}

\def\BibTeX{{\rm B\kern-.05em{\sc i\kern-.025em b}\kern-.08em
    T\kern-.1667em\lower.7ex\hbox{E}\kern-.125emX}}

\begin{document}

\title{High-efficiency, Wideband GRIN Lenses with Intrinsically Matched Unit-cells}

\author{Nicolas~Garcia,~\IEEEmembership{Student Member,~IEEE,}
        Jonathan~Chisum,~\IEEEmembership{Senior Member,~IEEE,}
\thanks{The authors are with the Department of Electrical Engineering, University of Notre Dame, Notre Dame, IN, 46556 USA (e-mail: ngarcia7@nd.edu; jchisum@nd.edu). Research was sponsored by Parry Labs and was accomplished under award \#PL2018-SB-001. Additional funding was provided by the NSF-sponsored BWAC I/UCRC under award number CNS-1439682-011.}
}

\maketitle

\begin{abstract} We present an automated design procedure for the rapid realization of wideband millimeter-wave lens antennas. The design method is based upon the creation of a library of matched unit-cells which comprise wideband impedance matching sections on either side of a phase-delaying core section. The phase accumulation and impedance match of each unit-cell is characterized over frequency and incident angle. The lens is divided into rings, each of which is assigned an optimal unit-cell based on incident angle and required local phase correction given that the lens must collimate the incident wavefront. A unit-cell library for a given realizable permittivity range, lens thickness, and unit-cell stack-up can be used to design a wide variety of flat wideband lenses for various diameters, feed elements, and focal distances. A demonstration GRIN lens antenna is designed, fabricated, and measured in both far-field and near-field chambers. The antenna functions as intended from 14\,GHz to 40\,GHz and is therefore suitable for all proposed 5G MMW bands, Ku- and Ka-band fixed satellite services. The use of intrinsically matched unit-cells results in aperture efficiency ranging from 31\% to 72\% over the 2.9:1 bandwidth which is the highest aperture efficiency demonstrated across such a wide operating band.
\end{abstract}





\section{Introduction}

\IEEEPARstart{T}{he last decade} has seen a tremendous growth in the global appetite for wireless microwave and MMW communications. The demand for ever-higher data capacity motivates the development of highly directive antenna platforms with beamforming capabilities. MMW and 5G beamforming transceivers are expected to become the status quo for various industries including cellular services\cite{Rappaport_2013, Hong_Baek_2014}, autonomous vehicles \cite{Kumari_2018, Menzel_Moebius_2012} and satellite communications. These small-cell applications in particular require highly directive antennas with low power consumption. Dielectric lens antennas are becoming increasingly viable in this space, exhibiting high directivity and even beam scan with minimal power consumption owing to their passive nature\cite{Imbert_2017, Fan_Yang_2018, Li_Ge_2019}. Flat dielectric lenses in particular are desirable for low-profile MMW and 5G applications due to their compact form factor and reduced weight. Flat lens antennas can also achieve low power beam-scan by means of a switched-feed network beneath the lens \cite{Bai_VTC2017,mateoFLLTO_APS, Imbert2014, Su_Chen_2018} or even a simple ultra-low-power radio at every feed location \cite{Gao_ITA_2017}. While flat lenses don't compare favorably to phased array systems in terms of scan loss\cite{Yang_2019, Kibaroglu_2018}, the high power requirements and overall cost of phased arrays may motivate the use of flat lenses in applications where aggressive scan angles are not required. 


\begin{figure}[t]
\centering
\includegraphics[width=3.5in]{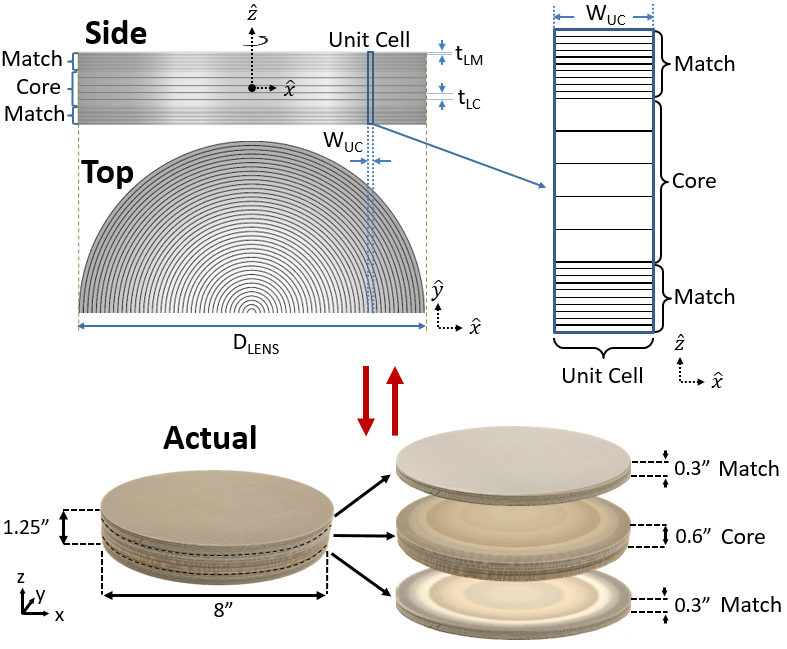}
\caption{Each lens has rotational symmetry about the central axis ($\hat{z}$-axis) and mirror symmetry across the center of the lens ($\hat{x}\hat{y}$-plane). The side-view shows the vertical stack-up comprised of a multilayer matching section on top and bottom surrounding a multilayer central core. Each layer in the matching and core sections have thickness $t_{LM}$ and $t_{LC}$, respectively. The top-view shows the rotational symmetry where the in-plane gradient is realized with rings, each of which is assigned a specific unit-cell and having width $W_{UC}$. Note that the bottom half of the ``top-view'' of the lens is cropped from the schematic for conciseness. Photos of the fabricated demonstration lens depict the decomposition of the lens into matching and core layers. The different colors in the layer cuts are due to the different substrates used to achieve in-plane gradients.}
\label{fig:lens_decomposition}
\end{figure}

In this work, the flat lens morphology of interest is the GRadient INdex (GRIN) lens. GRIN lenses achieve wave collimation by means of spatially varying index of refraction\cite{zhang_grintaper_2019, ChenCui_Broadband_2009, ChenCui_Broadband_2011, MahmoudKishk_2014, elef_matched_2017, elef_matched_2018}. Where homogeneous dielectric lenses are constrained in terms of geometry, a GRIN lens can theoretically achieve wave collimation with \textit{any} geometry given a sufficient range of refractive index (ie, an extreme enough \textit{GRIN profile})\cite{LeonhardtTO, PendryTO}. For flat GRIN lenses, this translates to lower form-factor/weight/cost via greater design freedom: lenses can be made thinner or with larger diameter and can be designed for smaller F/D. Flat lens GRIN profiles are often designed with either transformation optics \cite{LeonhardtTO,PendryTO,mateoFLLTO_APS,Bai_VTC2017}, quasi-conformal transformation optics (QCTO) \cite{LiQCTO}, or the field transformation (FT) method  \cite{MittraFT}. GRIN lenses are generally assumed to be axially symmetric and mirrored about the center of the lens along the bore-site axis. They typically comprise multiple stacked layers of dielectric with each layer having a unique in-plane permittivity gradient discretized into concentric rings. \textcolor{black}{The upper region of  Fig.\ref{fig:lens_decomposition} depicts a sample GRIN lens discretized into rings with constituent vertical layers. \st{These} The in-plane} gradients can be realized through various means including 3D printing \cite{Fan_Yang_2018, Xin_AM_2016} and perforated dielectrics \cite{PetosaPerforated,GarciaAPS}. \textcolor{black}{While 3D printing solutions offer immense freedom in artificial dielectric design, conventional RF-suitable 3D printable materials do not exceed $\epsilon_r$ = 3-4 \cite{acikgoz_mittra_3Dprint_2016, Fan_Yang_2018, Xin_AM_2016, Gbele_Liang_Ng_Gehm_Xin_2014}. Furthermore, while high-end 3D printer resolution specifications are often in the 10's of $\mu$m \cite{Xin_AM_2016} these are often not realizable or are prohibitively expensive, with typical 3D printed artificial dielectric cells being on the order of ~2-5\,mm \cite{acikgoz_mittra_3Dprint_2016, Fan_Yang_2018, Xin_AM_2016, Gbele_Liang_Ng_Gehm_Xin_2014}. In comparison, low loss RF PCB materials are available in a wide range of host dielectric constants ($\epsilon_r$ = 2-12) and commercial PCB milling companies can reliably fabricate thru-hole vias on the order of 100's of $\mu$m. For these reasons, perforated PCB substrates were chosen as the material basis for this work. \st{The upper region of  Fig.1 depicts a sample GRIN lens discretized into rings with constituent vertical layers.}}

In order to achieve high aperture efficiency lenses it is widely recognized that GRIN lenses must have an impedance match to the surrounding medium. This has often been realized by including a single matching layer in addition to the nominal GRIN profile design \cite{ChenCui_Broadband_2009,ChenCui_Broadband_2011,elef_matched_2017,elef_matched_2018}. In \cite{MittraFT} and \cite{MahmoudKishk_2014} multi-layer matching networks were used to achieve a low reflection across the diameter of the lens. Recently \cite{zhang_grintaper_2019} proposed a method whereby the entire lens index profile is designed with a built-in taper such that matching or anti-reflection coating did not have to be designed separately. However, they only discuss a linear index gradient and did not optimize the taper for wideband performance. Furthermore, the lens antenna in \cite{zhang_grintaper_2019} is only graded in one dimension and is intended to reside in the aperture of the feed horn antenna. 





In this article we present a library-based design process based on the FT method. The approach is similar to that of \cite{zhang_grintaper_2019} in that it addresses impedance match and phase collimation simultaneously but our process permits a wide variety of matching taper profiles and can cater to particular bandwidths while minimizing the overall lens thickness. The library comprises wideband impedance matched unit-cells which in turn comprise artificial dielectric structures --- the characteristics of which (maximum and minimum realizable index, lattice spacing relative to minimum wavelength, and overall geometry) ultimately constrain the library. The individual dielectrics are stacked vertically to make high permittivity phase-delay sections sandwiched between wideband impedance matching tapers. Fig. \ref{fig:lens_decomposition} details the overall lens structure with an inset showing the matching and core sections. The library is designed one time for a desired operating frequency band and material stackup (ie, number and thickness of layers in core and taper). The library-based approach has the advantage that a single wideband impedance matched library of unit-cells can be used for a variety of lens designs with different diameters, $F/D$, and even varying feed elements. Section\,\ref{sec:approach} details the design and fabrication of a matched unit-cell library (see Sec.\,\ref{sec:approach:library}) and how a lens is constructed from the library (see Sec.\,\ref{sec:approach:lens}) to achieve wideband, high aperture efficiency operation. An algorithm is presented to create an arbitrary diameter lens for any feed at any $F/D$. Sec.\,\ref{sec:demo} details the design of a lens to demonstrate the method and measurements are discussed in Sec.\,\ref{sec:results}. The wideband demonstration lens performance is then compared with the literature and, to the best of the authors' knowledge, is shown to exhibit state-of-the-art aperture efficiency over a wider bandwidth than any other result reported thus far.


\section{Approach} \label{sec:approach}

The lens design objective in this work is to achieve high aperture efficiency over a wide bandwidth. Aperture efficiency is unity for a uniform amplitude and uniform phase aperture field distribution. Therefore, for a given amplitude taper (typically set to achieve a prescribed sidelobe level), aperture efficiency is maximized by producing a flat phase distribution across the top of the lens.

Aperture efficiency is a useful metric because it includes all sources of gain reduction and, for our purposes, will be defined as, 
\begin{equation} \label{eq:ApEff}
    \textcolor{black}{\eta_{ap} = e_r\eta_{ill}|T\left(\theta_{inc},f\right)|\eta_a.}
\end{equation}
\noindent $e_r$ is the radiation efficiency which is approximately $1.0$ for GRIN lenses fabricated from low loss-tangent microwave substrates. $\eta_{ill}=\eta_t\eta_s$ is the illumination efficiency which is the product of the taper efficiency and the spillover efficiency.\textcolor{black}{ \st{For a given $F/D$ we estimate that a lens configuration with no amplitude shaping can achieve a maximum illumination efficiency of around $75$\%.$\eta_{ph}$ is the phase efficiency and is $1.0$ for a lens with perfect phase collimation.}} $|T\left(\theta_{inc},f\right)|$ is the magnitude of the angle- and frequency-dependent power transmission coefficient which considers any power lost to impedance mismatch. We will also refer to the normalized linear transmission coefficient as the transmission efficiency $\eta_T$. All other efficiencies are lumped into $\eta_a$, the achievement efficiency, and are not considered in this work since they can typically be nearly unity for a well-designed system \cite[Ch.~9.6]{Stutzman}.  \emph{Therefore, the design of a high aperture efficiency, wideband GRIN lens antenna is tantamount to achieving a high angle- and frequency-dependent transmission coefficient, $|T\left(\theta_{inc},f\right)|$.}


\subsection{Lens Composition}

In order to accomplish both flat phase and a wideband transmission efficiency, the GRIN lens must have an in-plane ($\hat{x}\hat{y}$) permittivity gradient (for phase collimation) and a beam-axis ($\hat{z}$) permittivity gradient (for impedance matching). Figure\,\ref{fig:lens_decomposition} shows the composition of such a lens whose beam-axis stackup is comprised of a matching region on top and bottom and a central core region in which the majority of the phase is accumulated. The radial phase profile is approximated by breaking the lens into concentric rings, each of which comprises a single unit-cell. The beam-axis stackup per unit-cell is shown in detail in the upper-right inset of Fig.\,\ref{fig:lens_decomposition}. Each unit-cell is matched over the desired band by incorporating the matching sections on the top and bottom of the beam-axis stackup. Matching sections are generally realized as slowly varying impedance tapers along the beam-axis and therefore, to reduce reflections due to discretization, the layer thickness in the matching sections, $t_{LM}$, is generally less than the layer thickness in the core, $t_{LC}$.

For such a construction the design objectives reduce to the following requirements for each unit-cell: i) the matching layers should maintain a prescribed return loss and insertion loss over the desired frequency band, ii) the core thickness should be maximized, and iii) the quantization amongst unit-cells in the library should be minimized.

\subsection{Wideband impedance matching}

The requirement to realize an impedance match over a wide bandwidth prefers thick matching sections. 
There are many classical taper profiles including the exponential, the triangular, and the Klopfenstein which generally provide various tradeoffs between frequency response (especially the frequency of the first reflection zero) and the in-band match performance.

\begin{figure}
    \centering
    \includegraphics[width=3in]{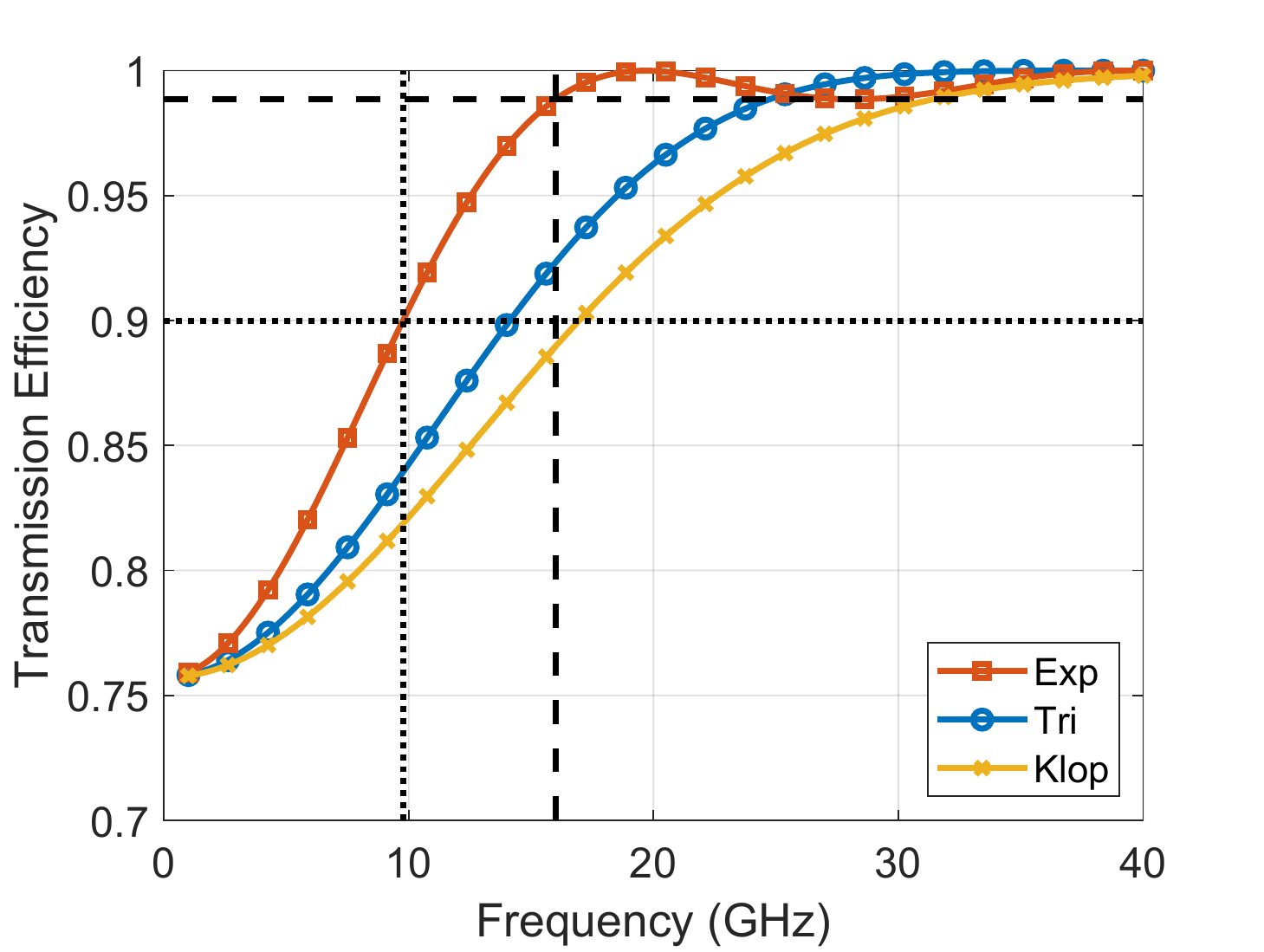}
    \caption{The maximum transmission efficiency for an exponential taper of length $0.300"$ from $10\,$ to beyond $40\,$GHz (high-pass match) is maintained above $90$\% for the exponential taper which permits thinner overall lenses than other tapers.}
    \label{fig:TaperComparison}
\end{figure}

Figure\,\ref{fig:TaperComparison} shows the transmission coefficient over frequency, $|T(f)|$, for the three tapers mentioned when the matching section is constrained to a thickness of $0.300"$ or $10$ layers of Rogers AD-series $0.03"$ boards. In each case the match is between free space ($\eta_0=377\Omega$) and the impedance of the highest permittivity unit-cell, $\eta_{UC} = \eta_0/\sqrt{\epsilon_r}=377/\sqrt{7.2}=137\Omega$. The taper dimensions and permittivity range are selected in anticipation of the demonstration lens described in Sections\,\ref{sec:approach:library} and \ref{sec:approach:lens}. The most obvious conclusion from Fig.\,\ref{fig:TaperComparison} is that the exponential taper provides the poorest match in the passband but it achieves an acceptable match (e.g., $90$\% transmission) at a much lower frequency than the triangular or Klopfenstein tapers. The thick dashed line shows the transmission coefficient at the first passband dip of the exponential taper (corresponds to the first reflection peak in the passband) at $0.9885$ and a frequency of $16$\,GHz\textcolor{black}{\st{---arguably an unnecessarily high transmission coefficient given the illumination efficiency peak of $75$\%}}. If the acceptable transmission efficiency level is deemed to be $90$\% (fine dashed line) the lowest operating frequency is $9.8$, $14$, and $17$\,GHz for the exponential, triangular, and Klopfenstein tapers, respectively. \textcolor{black}{\st{A lens with a $90$\% transmission efficiency and an optimal illumination efficiency around $75$\% will achieve a maximum aperture efficiency of $68$\%}}. For this work we have selected the exponential taper because, while the in-band transmission efficiency varies more than the others, the match band begins at a lower frequency for a given length. A thinner matching section allows the central core to comprise a higher percentage of the total lens thickness and the lens can be thinner for a given phase accumulation.

\subsection{Wideband Unit-cell Library} \label{sec:approach:library}

As stated above, each unit-cell is comprised of impedance matching tapers which match from the impedance of free space, $\eta_0=\sqrt{\frac{\mu_0}{\epsilon_0}}$, to the impedance of the unit-cell core, $\eta_{core}=\sqrt{\frac{\mu_0}{\epsilon_{core}\epsilon_0}}$. A library of matched unit-cells is then created by matching all possible core impedances on top and bottom to the impedance of free space by means of an exponential taper.

\begin{figure}
    \centering
    \includegraphics[width=3in]{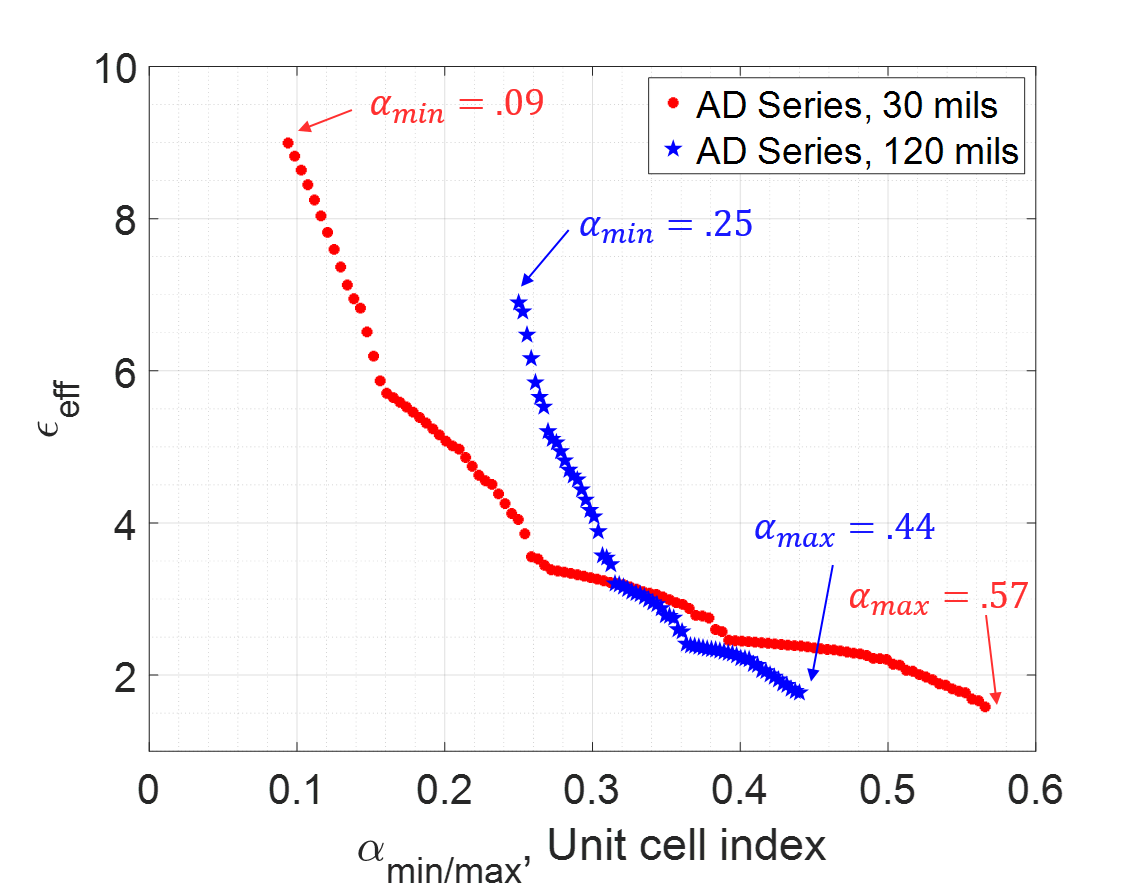}
    \caption{Permittivity list of perforated dielectrics in Rogers AD-Series substrates. The red circle trace corresponds to $30$\,mil substrates and the blue star trace corresponds to $120$\,mil substrates }
    \label{fig:permittivity_list}
\end{figure}

\begin{figure}
    \centering
    \includegraphics[width=3in]{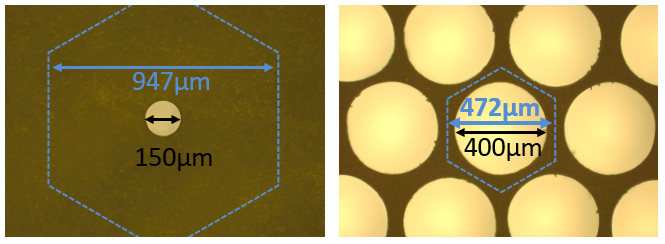}
    \caption{Backlit photos of drilled PCB substrates for most extreme fill factor $\alpha$ values at 5x magnification. Left: $150$\,$\mu$m hole in a $947$\,$\mu$m unit-cell in AD250 ($\epsilon_h$ = $2.5$) for $\alpha$ = $0.02$.  Right: $400$\,$\mu$m hole in a $472$\,$\mu$m unit-cell in AD1000 ($\epsilon_h$ = $10.35$) for $\alpha$ = $0.65$}
    \label{fig:pcb_closeup}
\end{figure}

The unit-cell library is constrained by the material platform that generates the various permittivity values. The library used in this article is based on perforated low-loss RF substrates in the Rogers AD-Series: AD1000, AD600, AD350 and AD250. These materials range in dielectric constant from $2.5$ to $10.35$ and our experiments indicate a range of $\epsilon_{eff}$ values from $1.54$ to $9.15$ are possible. Fig.\,\ref{fig:permittivity_list} shows the experimentally determined discrete values of all realizable effective permittivity values which can be employed in the unit-cells (both taper and core sections). This list includes values for $0.030$\,inch (``$30$\,mil'') and $0.120$\,inch (``$120$\,mil'') substrates. The thinner substrates are generally used in the matching layers (i.e., $t_{LM}=30$\,mils) to provide finer discretization while the thicker substrates are generally used in the core to reduce overall lens cost (i.e., fewer total drills are required in lenses with thicker substrates). To exploit the full permittivity range the lens comprises all substrates so a given layer may have concentric rings of different substrates.
Figure \ref{fig:pcb_closeup} shows 5x magnification of perforated dielectrics in AD Series AD250 (left) and AD1000 (right). The perforations are arranged on a hexagonal lattice whose lattice constant $L_C$ is determined by the background permittivity of the host substrate, $\epsilon_h$. \textcolor{black}{The lattice constant is constrained such that $L_C/\lambda_{\mathrm{guided, host}}$ = 0.2 for the highest operating frequency.} The $\epsilon_h$ of AD1000 is substantially larger than that of AD250 and corresponds to a much smaller $L_C$ --- $472$\,$\mu$m in AD1000 compared to $947$\,$\mu$m in AD250\textcolor{black}{. These values correspond to $L_C  = [\lambda_0/16_{\mathrm{AD1000}}, \lambda_0/8_{\mathrm{AD250}}]$ at 40$\,$GHz and are conservative compared to \cite{imbert_design_2015_perf_lens, petosa_design_2003, he_matched_2018} while enabling a reasonable range of commercially available drill sizes. \st{ --- to guarantee that a given $L_C$ is substantially smaller than the corresponding material's guided wavelength at $40$\,GHz}}. Note that because the substrate $L_C$ values are generally much smaller than $W_{UC}$, a given layer in a unit-cell in the eventual lens would comprise many identical perforations. The effective permittivity of a given perforated hexagon is determined by $\epsilon_h$ and the fill factor $\alpha$: the volumetric ratio of the perforation volume to the total 3D hexagonal-prism volume. The range of $\epsilon_{eff}$ for a substrate is then constrained by the range of drill diameters that can be used in that substrate's unique $L_C$. 
Table \ref{tab:pcb_table} lists the material specifications for the $120$\,mil substrates in Fig.\,\ref{fig:permittivity_list}: $\epsilon_h$, $L_C$, maximum drill diameter (all materials used a minimum drill diameter of $0.25$\,mm), $\alpha$-range, and $\epsilon_{eff}$. Clearly the $30$\,mil substrates can express a greater range of $\epsilon_{eff}$ --- as indicated by Fig.\,\ref{fig:permittivity_list} --- but the lens uses both substrate thicknesses and requires that they express the same permittivity range to maintain smooth transitions at the taper-core interface. For this reason the usable permittivity range of the lens is that of the $120$\,mil material: $1.67$ to $7.18$. Only these values are considered for the lens unit-cell library moving forward.


\begin{table}
\centering
\caption{Material Specifications for PCB Artificial Dielectrics Comprising unit-cell Library}
\begin{tabular}{@{}r c c c c c@{}}
\toprule
    Material & $\epsilon_h$ & $L_C$ (mm) & Max Drill (mm)  & $\alpha$-range & $\epsilon_{eff}$-range\\ 
\cmidrule{1-6}
    AD1000 & 10.20 & .47 & 0.32 & 0.26-0.42 & 7.18-5.57 \\
    AD600 & 6.30  & .60 & 0.41 & 0.16-0.42 & 5.21-3.65 \\
    AD350 &  3.50   & .80 & 0.50 & 0.09-0.35 & 3.22-2.45\\ 
    AD250 & 2.52   & .95 &  0.65 & 0.06-0.42 & 2.39-1.78\\ 
\bottomrule
\end{tabular}
\vspace{2mm}

\vspace{-7mm}
\label{tab:pcb_table}
\end{table}

\begin{figure}
    \centering
    \includegraphics[width=3.5in]{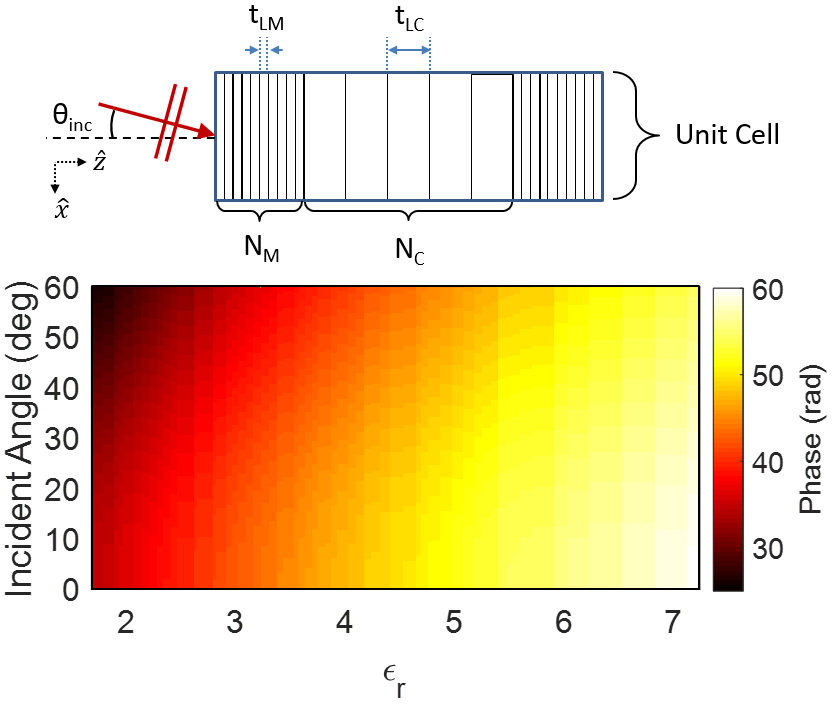}
    \caption{Non-normal incidence on unit-cell (top). Phase accumulated for all unit-cells (denoted by their maximum permittivity) for various incident angles (bottom) at $40$\,GHz.}
    \label{fig:taper_phase_epsilon}
\end{figure}

\begin{figure}
    \centering
    \includegraphics[width=3.5in]{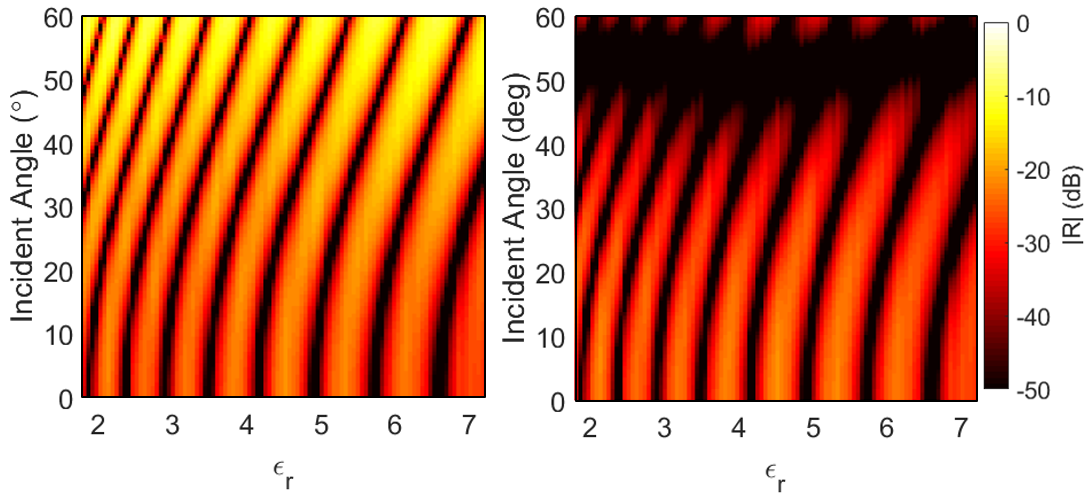}
    \caption{Return loss for TE (left) and TM (right) modes for  all unit-cells (denoted by their maximum permittivity) shown at various incident angles at $40$\,GHz.}
    \label{fig:taper_ref_epsilon}
\end{figure}

\begin{figure}
    \centering
    \includegraphics[width=3.5in]{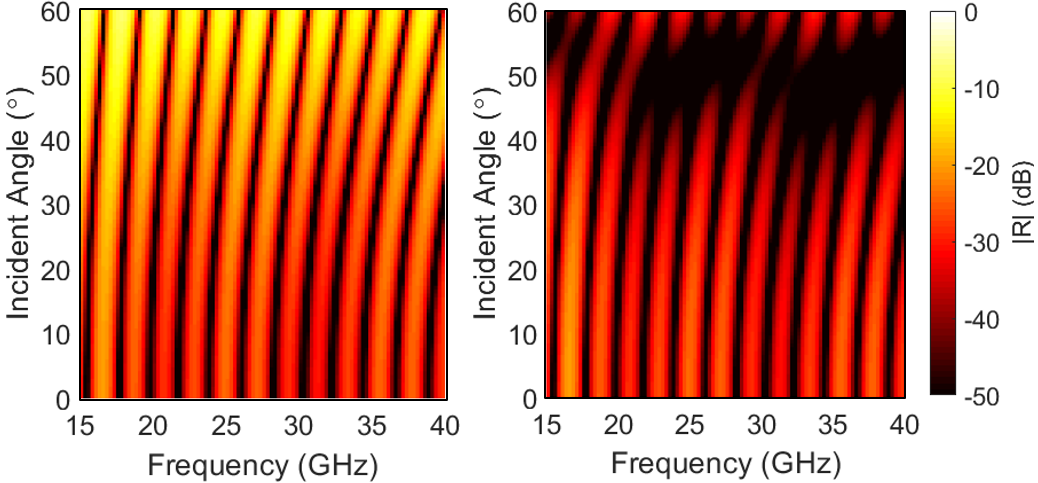}
    \caption{Return loss for TE (left) and TM (right) modes over frequency given the highest core permittivity unit-cell ($\epsilon_{eff}=7.18$). }
    \label{fig:taper_ref_frequency}
\end{figure}

By matching with exponential tapers from free-space to all possible core permittivity values a library of all possible unit-cells is generated. Each matched unit-cell is indexed by the maximum permittivity contained in the core of the unit-cell and is characterized by the total accumulated phase versus incidence angle. A ray tracing approximation is used to analyze each unit-cell for a plane-wave incident from $0$--$60^\circ$ and impedance match and phase accumulation are computed from Fresnel coefficients and Snell's law. Referring to the inset schematic in Fig.\,\ref{fig:taper_phase_epsilon}, the total phase of each unit-cell is computed as,
\begin{equation}
    \phi[idx] = 2\sum_{n=1}^{N_m}\beta[n]t_{LM}+\sum_{n=N_m+1}^{N_m+N_c}\beta[n]t_{LC},
\end{equation}
\noindent where $N_m$ is the total number of layers in each matching section with layer thickness $t_{LM}$, and $N_c$ is the total number of layers in the core with layer thickness $t_{LC}$. 

Fig.\,\ref{fig:taper_phase_epsilon} shows the total accumulated phase of each unit-cell at $40\,$GHz over all incidence angles. Since the phase accumulated is a function of incidence angle the radial distance to each unit-cell in a lens cross-section is critical to selecting the proper phase accumulation at that point. Each unit-cell of this library is comprised of ten, $30$\,mil layers in each matching section (top and bottom) and five $120$\,mil layers in the core. The library can achieve a total phase differential of approximately $30\,$radians ($58\,$rad$-28\,$rad). Figure\,\ref{fig:taper_ref_epsilon}  and Figure\,\ref{fig:taper_ref_frequency} show the return loss of TE-modes and TM-modes for various incidence angles over permittivity index and frequency of operation, respectively; Fig.\,\ref{fig:taper_ref_frequency} assumes the highest core permittivity.

\begin{figure}
    \centering
    \includegraphics[width=3.5in]{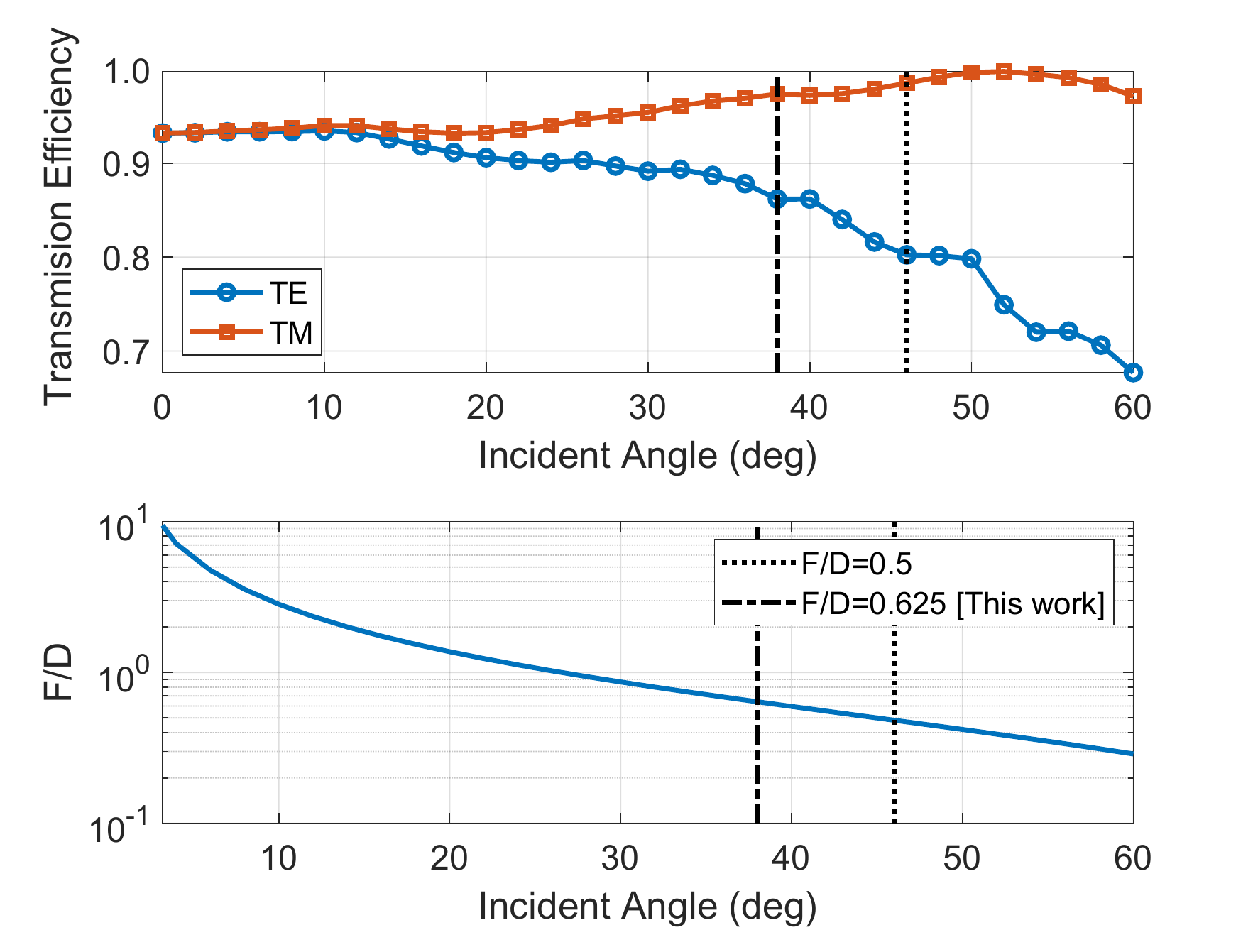}
    \caption{(top) Worst-case TE- and TM-mode transmission coefficients at $40\,$GHz for all unit-cells in the library (Fig.\,\ref{fig:taper_phase_epsilon}) versus incident angle out to $60^\circ$. (bottom) F/D for a given maximum angle (of the marginal ray) incident upon the lens from the feed. For $F/D=0.5$ the maximum angle is $46^\circ$ with TE- and TM-mode transmission efficiency better than $0.932$ and $0.803$, respectively. For $F/D=0.625$ (used in the demonstration lens in this work) the maximum angle is $38^\circ$ with TE- and TM-mode transmission efficiency better than $0.862$ and $0.932$, respectively.} 
    \label{fig:FresnelTransmission}
\end{figure}

The worst case TE- and TM-mode transmission efficiencies for the whole unit-cell library are computed at $40\,$GHz and shown in Fig.\,\ref{fig:FresnelTransmission}. As is well known, as incidence angle increases the transmission coefficients for TE- and TM-modes diverge, eventually leading to reduced transmission of TE-modes and perfect transmission of TM-modes at Brewster's angle. As such, in order for a lens to minimally perturb the polarization of the feed, the match must have sufficiently high transmission across both modes to maintain a good match across all angles. As shown, the worst-case TM-mode transmission efficiency is greater than $93$\% over all incidence angles of interest but the TE-mode transmission efficiency degrades at higher incidence angles. The worst-case TE-mode transmission efficiency is better than $90$\% out to $26^\circ$ and better than $86$\% out to $40^\circ$ so polarization will be minimally perturbed. For a typical $F/D$ range of $0.5$--$2.0$ incident angles are bounded to less than $46^\circ$ and the worst-case transmission efficiency for either polarization is $80.3\%$--$92.6\%$.

\subsection{Design Algorithm} \label{sec:approach:lens}
\begin{figure}
    \centering
    \includegraphics[width=3in]{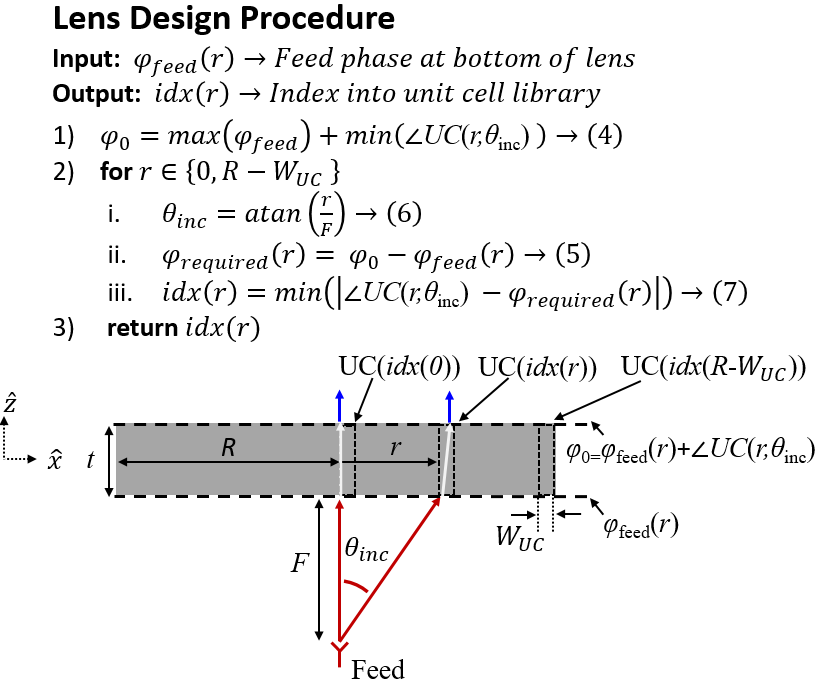}
    \caption{A lens of radius $R$ and thickness $t$ is comprised of concentric rings which are axially symmetric about the $z-$axis. Each ring is of width $W_{UC}$ such that there are $N=D/2W_{UC}$ rings. The cross-section of each ring is a unit-cell drawn from the wideband, wide-angle unit-cell library and consists of matching layers on bottom and top of a core. The total phase of each unit-cell at radial distance $r$, should be such that when added to the phase of the feed incident upon the bottom of the lens, $\phi_{feed}(r)$, the total accumulated phase is uniform across the top of the lens and equal to $\phi_0$.}
    \label{fig:algorithm}
\end{figure}

For a given feed element a lens is designed using the unit-cells from the wideband library. Figure\,\ref{fig:algorithm} shows the cross section of a lens of radius, $R$, overall thickness, $t$, unit-cell width, $W_{UC}$, and feed element focal distance, $F$, defined here as the distance from the phase-center of the feed element to the bottom surface of the lens \textcolor{black}{(shown in Fig. \,\ref{fig:algorithm})}. The origin lies in the center of the lens with the $z-$axis pointing toward the top of the lens and in the direction of wave propagation in the transmit mode. The lens is defined as rotationally symmetric about the $\hat{z}-$axis so it is fully defined in the $\hat{r}\hat{z}$-axis. The lens has both a vertical permittivity gradient (along the $\hat{z}-$axis) and a radial permittivity gradient (along the $\hat{r}-$axis). Since the lens is comprised of unit-cells the vertical gradient is defined from the unit-cell library created in the previous section. The task of lens design then becomes assigning unit-cells to particular radial locations along the $\hat{r}-$axis. This design approach assumes that incident fields are locally plane, even for a unit-cell of relatively small cross section (not electrically large). As such, the GRIN profile must be a smoothly varying profile such that if a ``ray'' would propagate from one unit-cell and into an adjacent unit-cell, the permittivity profile would not change significantly. 

In Fig.\,\ref{fig:algorithm} three unit-cells are indicated with dashed black lines--the first unit-cell to the right of the central axis of the lens at $r=0$ is labeled UC$(idx(0))$ and the last unit-cell to the right of the central axis of the lens at $r=R-W_{UC}$ is labeled UC$(idx(R-W_{UC}))$. For an arbitrary unit-cell at radius $r$, here labeled UC$(idx(r))$, the UC$(\cdot)$ indicates a particular unit-cell defined as a vertical stackup including a bottom and top matching layer and a core and characterized in the unit-cell library above. Since the lens is rotationally symmetric about the $\hat{z}-$axis each unit-cell actually describes the cross-section of a ring and the lens is composed of$N=D/2W_{UC}$ rings. The library above defines a collection of unit-cells specified by an index number therefore the GRIN lens design is reduced to creating a mapping $idx(r)$ which converts a radial distance to a unit-cell index number. Once the radial index vector is defined the $\hat{r}\hat{z}-$plane permittivity gradient can be drawn and the lens can be modeled and fabricated.

\textcolor{black}{Some care must be taken when assigning $W_{UC}$. Given an $r$ coordinate, a single unit-cell is characterized for phase accumulation and return loss at the corresponding incident angle. For large $W_{UC}$ there is substantial variation in incident angle and the phase accumulation across the unit-cell will vary in turn (see Fig.\,\ref{fig:taper_phase_epsilon}). This results in aperture phase perturbations which are periodic with $W_{UC}$. If $W_{UC}$ is larger than $\lambda_0$, these perturbations will result in sidelobes in the far field at $\theta=\sin^{-1}(\lambda_0/W_{UC}$)\cite{Brown_1950}. For this reason, $W_{UC}$ should not be larger than the wavelength of the highest operating frequency of the lens.}

In order to design a lens the feed element must be specified \textcolor{black}{insofar as its power pattern and focal distance. These parameters together inform the illumination efficiency of the lens but the focal distance in particular determines the phase collimating characteristic of the lens -- assuming it is in the far field of the feed. In this case the incident phase can be approximated by assuming an idealized isotropic source at the focal distance and, from \cite[Ch.~1.2]{Sletten}, the feed's field pattern can be modeled as, \st{because assuming an idealized isotropic feed to compute the incident phase of the feed, $\phi_{feed}$, is not sufficient to achieve high aperture efficiency lens antennas. Instead, a specific feed element is either electromagnetically simulated or measured to determine the phase of the field incident upon the bottom surface of the lens,}}

\begin{equation}\label{eq:feedPattern}
        \textcolor{black}{G(\theta)\approx A\,\cos^N(\theta) \left[ \sin(\phi)\boldsymbol{\hat{\theta}} + \cos(\phi)\boldsymbol{\hat{\phi}}\right]e^{-j \beta \rho}/\rho} 
    \end{equation} 
\noindent \textcolor{black}{where the $e^{-j\beta \rho}/\rho$ term models an outwardly expanding spherical wave originating from a point source at the feed phase center ($\rho$ is the radial distance from the phase center). The term in $\left[\cdot\right]$ describes the angle-dependent polarization for a y-polarized feed. $A\cos^N(\theta)$ describes the amplitude taper of the feed. The resulting phase at the bottom of the lens is}

\begin{equation}
    \phi_{feed} = \angle \vec{E}(x,y=0,z=-t/2).
\end{equation}
\noindent The lens must equalize this phase to a constant value $\phi_0$, where
\begin{equation} \label{eq:phi0}
    \phi_0 = max(\phi_{feed})+min(\angle UC),    
\end{equation}
\noindent where $max(\phi_{feed})$ is the maximum phase incident upon the bottom of the lens and $min(\angle UC)$ is the minimum phase of all unit-cells in the unit-cell library. The phase accumulated in free-space between the feed element and the bottom surface of the lens, plus the phase accumulated in the lens is constant over the radius of the lens and equal to (\ref{eq:phi0}). Therefore, 
\begin{equation} \label{eq:phiRequired}
    \angle UC = \phi_0-\phi_{feed},
\end{equation}
\noindent which is the phase that must be provided by a particular unit-cell in the library.

The design procedure is described in Fig. \ref{fig:algorithm}. First, compute the phase which the lens must accumulate in order to equalize the phase at the top of the lens from (\ref{eq:phi0}). This is an offset version of the inverse of the feed phase profile at the bottom of the lens. Then, for each radial location across the lens, $r \in \{0,D/2-W_{UC}\}$, compute the incident angle from the center of the lens based upon the geometric ray approximation,
\begin{equation} \label{eq:thetaInc}
    \theta_{inc} = \tan^{-1}\left(\frac{r}{F}\right).
\end{equation}
\noindent At this radial distance compute the phase required from the lens unit-cell using (\ref{eq:phiRequired}). Then look up the unit-cell in the library which has a total accumulated phase at $\theta_{inc}$ which most closely matches the required phase, $\phi_{required}$: 
\begin{equation} \label{eq:RingIdx}
    idx(r) = argmin(| \angle UC(r,\theta_{inc})-\phi_{required}(r) |).
\end{equation}
\noindent Assign this unit-cell's $\hat{z}-$axis permittivity profile to the cross-section of the ring at the radial location $r$. Note that since the entire library was designed to ensure a wideband, wide-angle impedance match, the only criterion for unit-cell selection is the total phase accumulated in the cell. Once the index into the unit-cell library is defined for each radial distance from $0$ to $D/2-W_{UC}$, the entire lens permittivity profile is known.

\subsection{Modifying a Lens Design}

Following the design algorithm in the previous section it is a simple matter to design a wide range of lenses based on a single unit-cell library. For example, a unit-cell library which exhibits a good impedance match over a desired frequency range and possesses a particular phase difference across the library can be used to design a wide variety of lenses with different diameters and different focal distances. As discussed in Sec.\,\ref{sec:approach:library}, lenses with an $F/D$ from $0.5$--$2.0$ (or larger if desired) and a transmission efficiency greater than or equal to $90$\% can be rapidly synthesized with the designed library. Similarly, a feed element can be exchanged and the lens redesigned by following the procedure in Fig.\,\ref{fig:algorithm} for the new gain pattern (and hence new optimal $F/D$) and feed phase center.

In this sense, once the matched library is created additional lenses can be rapidly synthesized and will immediately inherit a good impedance match. Of course if the diameter of a lens is increased significantly or the focal distance is reduced beyond the allowed range the total phase difference across the bottom of the lens may increase beyond that supported by the library and this may require a thicker overall lens to provide the necessary phase differential to equalize the phase at the top of the lens antenna. In this case a new library would be necessary but it will often be sufficient to simply thicken the core and not modify the matching sections.

\section{Demonstration} \label{sec:demo}

\subsection{Lens Description}

In order to demonstrate the proposed matched-library design approach, we designed an 8\," diameter lens from a unit-cell library corresponding to the perforated low loss RF substrates in Fig.\,\ref{fig:permittivity_list} . The input amplitude and phase profile used were those of a standard gain WR-28 $10$\,dBi horn antenna simulated in HFSS full-wave electromagnetic simulation software. This particular band was chosen because it contains both the $28$\,GHz and $39$\,GHz 5G MMW bands and because the artificial dielectric materials in our unit-cell library were designed for operation up to $40$\,GHz. With Empire XPU FDTD full-wave electromagnetic software, we simulated various permittivity profiles corresponding to different F/D values when fed with the WR-28 horn. We settled on a focal distance of $5$\," corresponding to F/D of $0.625$. 

\begin{figure}
    \centering
    \includegraphics[width=3.5in]{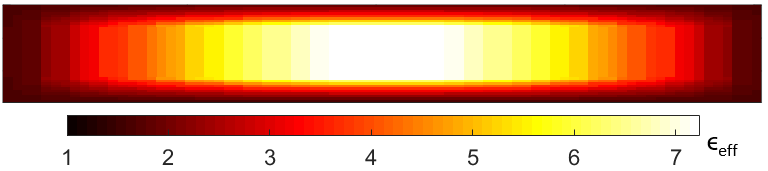}
    \caption{Lens permittivity profile, varying from $\epsilon_{eff}$ = $1.67$ at the edges to $\epsilon_{eff}$ = $7.18$ in the center.}
    \label{fig:permittivity_heatmap}
\end{figure}

As stated in Sec.\,\ref{sec:approach}, $30$\,mil substrates were used in the taper sections and $120$\,mil substrates were employed in the core. For this diameter and $F/D$ the total phase difference necessary could be realized with five $120$\,mil substrates in the core section and ten $30$\,mil substrates in each taper. This resulted in a nominal thickness of $1.2$\,". The GRIN profile resulting from the procedure in Fig.\,\ref{fig:algorithm} is shown in Fig.\,\ref{fig:permittivity_heatmap} where the highest effective permittivity employed was $7.18$ (in the radial center of the core) and the lowest effective permittivity was $1.67$ (at the bottom and edges of the lens). The lens layers and constituent substrates were etched and drilled by a commercial RF PCB foundry. All layers were then bonded together. The outward facing layers in the matching sections were entirely realized in AD250 to achieve the lowest effective permittivity while the inward facing match layers and all core layers comprise all four AD-series substrates to achieve the widest permittivity range (this is shown in the bottom right inset of Fig.\ref{fig:lens_decomposition}). Fully assembled the lens is $1.25$\," thick -- $50$\,mils thicker than nominal. This is likely due to the tolerance in the thickness of the Rogers substrates, where $120$\,mil substrates can vary up to $6$\,mils. The demonstration lens is $1.63$\,kg ($3.6$\,lbs) for an average density of $1.6$ \,g/cm$^3$. Total hole count was $1.6$ million. Figure\,\ref{fig:lens_decomposition} shows a photograph of the final bonded lens as well as an exploded view to detail the top and bottom matching layers and the phase accumulating core, with nominal thicknesses of $0.3"$ and $0.6"$, respectively.



\subsection{Impedance Response}
\begin{figure}
    \centering
    \includegraphics[width=3.5in]{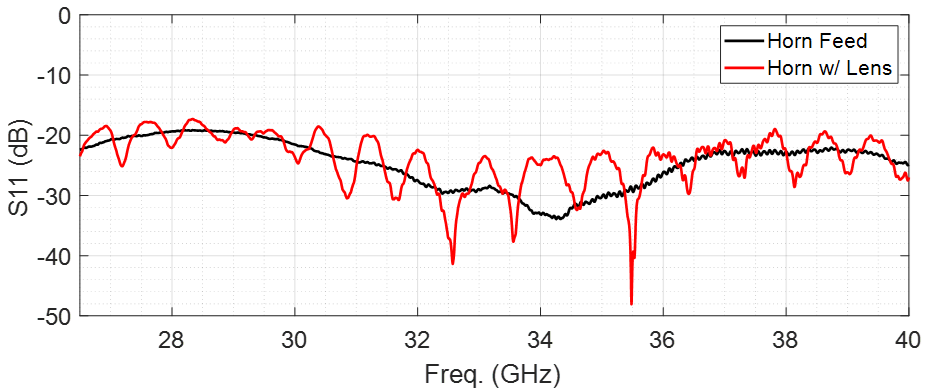}
    \caption{\textcolor{black}{S11 of the feed horn (black trace) and the feed horn with lens (red trace). }}
    \label{fig:sparams}
\end{figure}

\textcolor{black}{The calibrated S11 for the antenna system (feed horn and lens) in the WR-28 band (26.5-40\,GHz) is shown in Fig. \ref{fig:sparams}. The return loss of the lens system tracks the return loss of the horn with a small ripple across the band. The periodicity of this standing wave is roughly $840$\,MHz -- this corresponds to a freespace propagation distance of $35.7$\,cm and indicates an electrical discontinuity roughly $17.8$\,cm from the calibrated phase reference. This is nearly the path length from the WR-28 calibration reference to the lens: $15.2$\,cm (the focal distance to the feed is $12.7$\,cm and the horn length is approximately $2.5$\,cm). Considering that the lens does not have well-defined reflective boundaries, some amount of the lens thickness provides the remainder of the path length ($~2.5$\,cm). The system return loss is below 20\,dB across nearly the whole band, indicating that ripple is very mild and there is satisfactory impedance match at the WR-28 waveguide/horn interface. }

\textcolor{black}{However, because this measurement only considers the portion of the lens-reflected power that is incident on the horn antenna, it does not address spreading loss and is not a good indicator of the overall transmission efficiency. The S11 values in Fig. \ref{fig:sparams} correspond to reflections off a small radial extent of the lens and a limited range of incidence angles. Because the incident angle and unit-cell vary with radius, this S11 value is not representative of the whole lens. The effective transmission efficiency of the whole lens must be calculated based on the far field of the lens and other efficiencies; this analysis is provided in Sec.\,\ref{sec:demo}-D.}

\subsection{Pattern Measurements} \label{sec:results}

\begin{figure}
    \centering
    \includegraphics[width=3.5in]{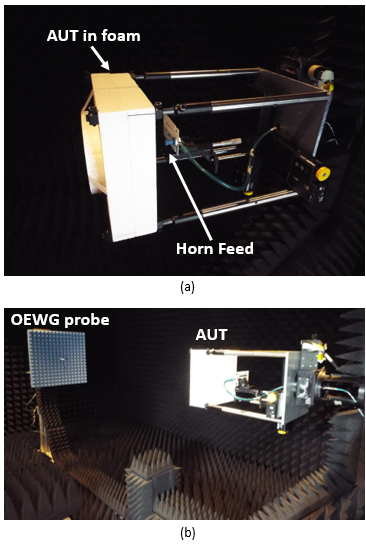}
    \caption{\textcolor{black}{Photos of the near field lens measurement setup: a. lens (encased in low dielectric constant foam) and feed with associated alignment hardware, b. lens system in anechoic near field chamber with open ended waveguide probe.}}
    \label{fig:nearfieldPhotos}
\end{figure}

\begin{figure}
    \centering
    \includegraphics[width=3.5in]{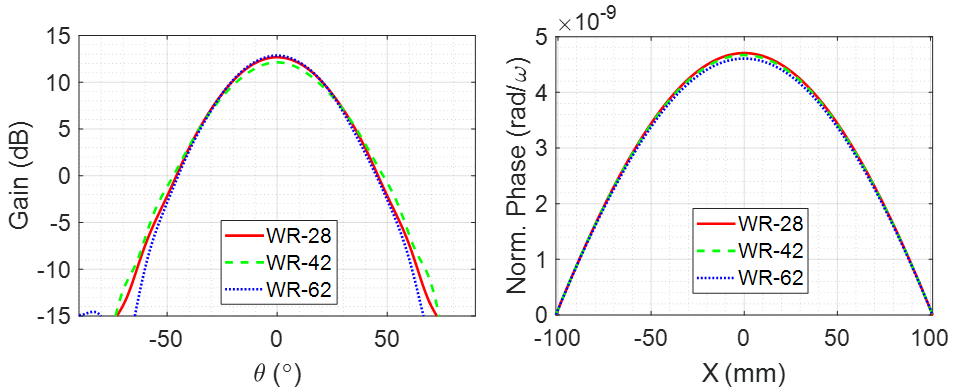}
    \caption{\textcolor{black}{Comparison of WR-28 (red solid trace), WR-42 (green dashed trace), and WR-62 (blue dotted trace) 10dBi horn characteristics: gain pattern (left) and normalized phase incident on lens surface (right). All cuts are taken in the E-plane and correspond to the highest operating frequency for each waveguide band: 40\,GHz, 26.5\,GHz, and 18\,GHz for WR-28, WR-42, and WR-62, respectively.}}
    \label{fig:feedComparisonSmall}
\end{figure}

\begin{figure}
    \centering
    \includegraphics[width=3.5in]{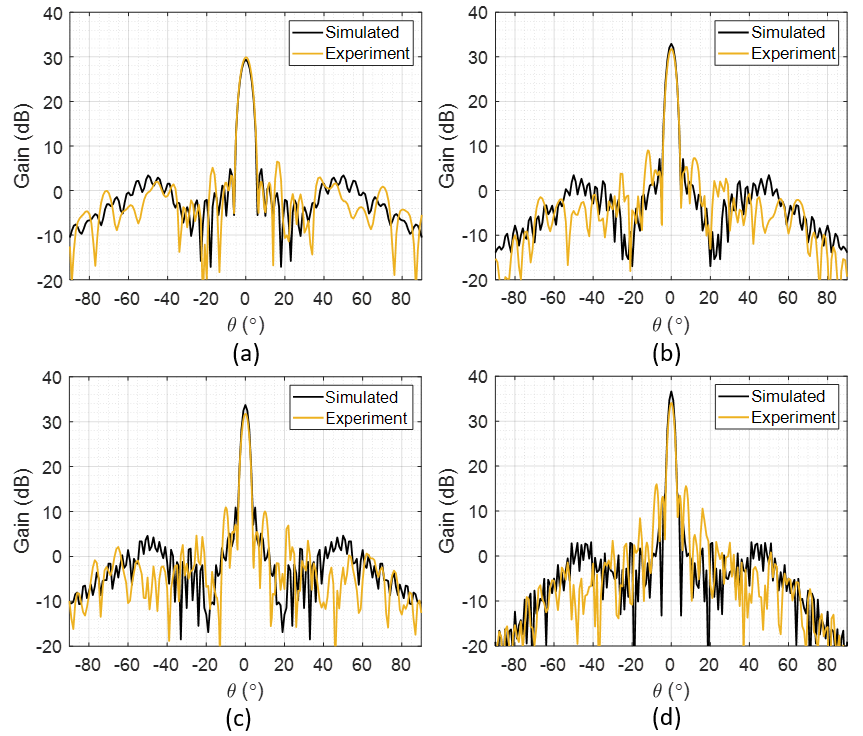}
    \caption{Simulated and measured far-field gain patterns at $18$ (a), $25$ (b), $30$ (c), and $40$\,GHz (d).}
    \label{fig:gain_patterns}
\end{figure}

Antenna measurements were taken in a far field chamber using the gain substitution method \cite[Ch.~13.3]{Stutzman}. Since the far-field of this antenna is greater than $30$\,feet --- significantly larger than the available anechoic chamber --- a Keysight N5225A network analyzer with time-gated S-parameters was used to remove errors due to multipath. \textcolor{black}{Measurements were also taken on a spherical near field range in an anechoic chamber. Photos of the lens system and near field testing environment are shown in Fig.\,\ref{fig:nearfieldPhotos}. The lens is encased in low dielectric constant foam and supported on struts as seen in Fig.\,\ref{fig:nearfieldPhotos}(a) The near field anechoic chamber and associated near field probe are shown in Fig.\,\ref{fig:nearfieldPhotos}(b).} Initially the lens was characterized from $26.5$\,GHz to $40$\,GHz with a $10$\,dBi WR-28 horn antenna. Subsequent measurements used $10$\,dBi WR-42 and WR-62 horn antennas to characterize the lens performance from $18$\,GHz to $26.5$\,GHz and $12$\,GHz to $18$\,GHz, respectively. \textcolor{black}{Fig.\,\ref{fig:feedComparisonSmall} (left) shows the E-plane gain patterns of the three feeds. Considering that the feed patterns are extremely similar and the lens is in the far field of each feed, this indicates all feeds will have similar edge-taper and spillover characteristics and are well represented by (\ref{eq:feedPattern}). Fig.\,\ref{fig:feedComparisonSmall} (right) shows the the normalized phase incident on the lens surface from each feed at its highest frequency of operation. The phase plots are normalized by angular frequency $\omega$ to better compare phase accumulation (note that the phase of the feed model given in (\ref{eq:feedPattern}) varies linearly with frequency). Because the lens is a true time delay device and the incident normalized phase distributions in Fig.\,\ref{fig:feedComparisonSmall} (right) are extremely similar, we expect performance to be similar for all feeds given that they are at the correct focal distance.
\st{Considering that both the WR-28,  WR-42, and WR-62 horn antennas have similar gain patterns over the respective frequencies, we expected both setups to have similar edge-taper.}} Furthermore, any difference in relative feed phase center could be accounted for by slightly shifting the focal distance of a given horn. Simulated and experimental gain patterns \textcolor{black}{of the lens system} at $18$\,GHz, $25$\,GHz (WR-42 frequencies), $30$\,GHz and $40$\,GHz (WR-28 frequencies) are given in Fig.\,\ref{fig:gain_patterns}. A detailed discussion of the gain and aperture efficiency is deferred until Sec.\,\ref{sec:Wideband} where gain and efficiency over a wide operating bandwidth is considered.

Generally speaking there is good agreement between simulated and experimental patterns. The notable discrepancies are the unexpected drop in experimental gain in the WR-28 band and the sudden appearance of high close-in sidelobes in all experimental results. Understanding the source of these discrepancies begins with recognizing approximations made in the simulated results: this lens is an electrically massive structure -- with a diameter of $27\,\lambda_0$ at $40$\,GHz -- comprising some $1.6$ million sub-$\lambda_0$ cells. Our simulations therefore do not consider each perforation (as this would be computationally intractable) but only consider ringed layers of homogeneous material as given by the unit-cell library. While this approximation adequately captures the permittivity gradient present in the real lens it also smooths over sharp dielectric discontinuities due both to rapidly changing hole sizes and substrate junctions in the artificial dielectric media. Considering that the lens is axially symmetric, we expect these errors to appear as periodic perturbations in the lens aperture field distribution. 

\begin{figure}
    \centering
    \includegraphics[width=3.5in]{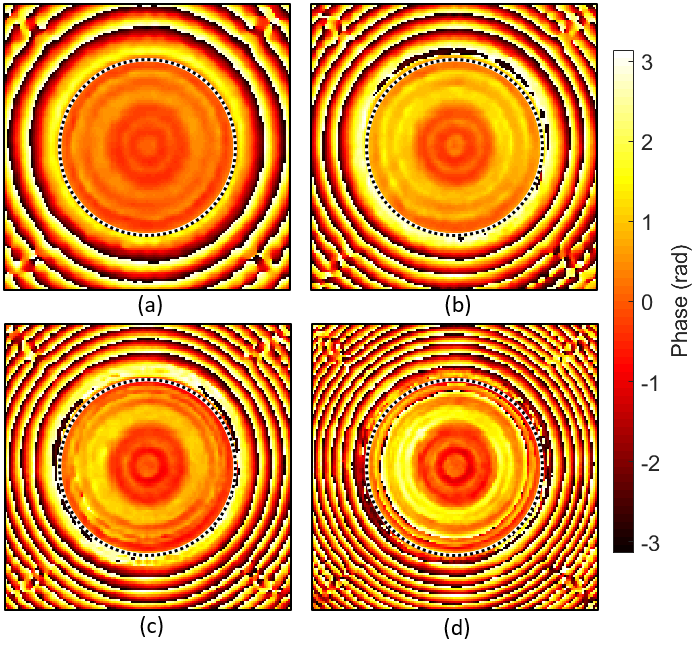}
    \caption{Aperture phase for dominant pol: a. $18$\,GHz, b. $25$\,GHz, c. $30$\,GHz, d. $40$\,GHz}
    \label{fig:aperture_phase}
\end{figure}

\begin{figure}
    \centering
    \includegraphics[width=3.5in]{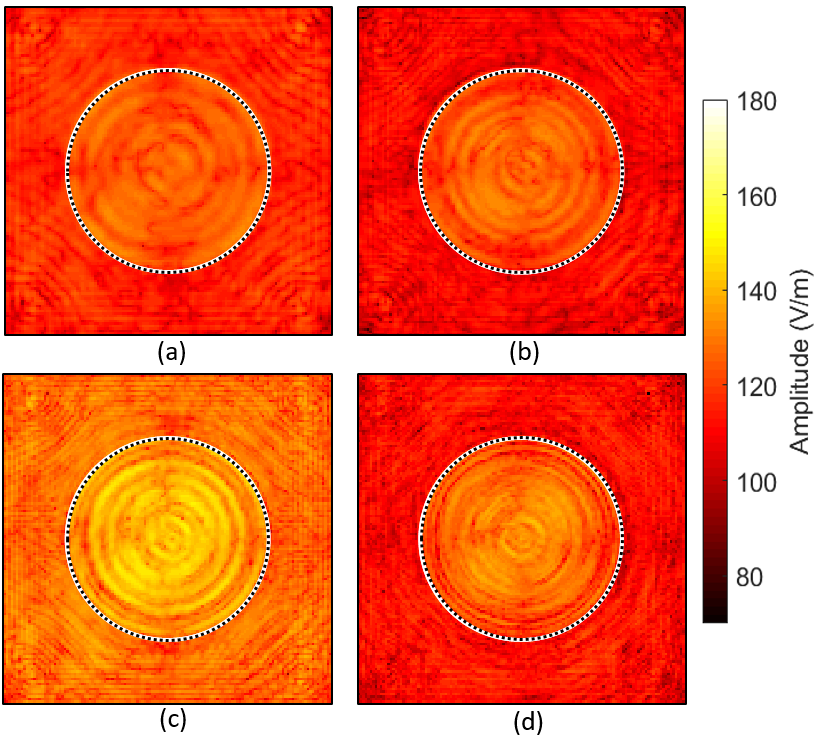}
    \caption{Aperture amplitude for dominant pol:  a. $18$\,GHz, b. $25$\,GHz, c. $30$\,GHz, d. $40$\,GHz. Perturbations at the corners of the images are due to slight interference of conductive rods used to support the lens during the near field measurement. }
    \label{fig:aperture_amplitude}
\end{figure}

In order to identify any periodic perturbations (and more generally assess the fabrication quality of the lens), the demonstration lens\textcolor{black}{' aperture fields were \st{was tested in a near field radiation chamber and aperture fields}} extracted via holographic back projection of the near field measurements. The absolute magnitude of the aperture and phase of the dominant polarization are shown for $18$\,GHz, $25$\,GHz, $30$\,GHz, and $40$\,GHz in Fig.\, \ref{fig:aperture_phase} and Fig.\,\ref{fig:aperture_amplitude}.

It is apparent from both the amplitude and phase distributions that there are radially periodic perturbations that are seemingly independent of frequency, indicating that there are structural nonidealities within the lens. Also considering that the periodic perturbations exhibit axial symmetry, the culprit is not likely a localized fabrication flaw. We suspected that the arrangement of substrates within the lens or potentially rapid discontinuities in effective permittivity were responsible for these perturbations. While the concentric configuration of substrates in the taper vary layer to layer (and any modal perturbations due to material discontinuities would be evanescent in more homogeneous regions on either side) the core region of the lens has the same material arrangement throughout and makes up most of the thickness of the lens. Indeed, the substrate junctions in the core line up nearly perfectly with the aperture perturbations in Fig.\,\ref{fig:aperture_phase} and Fig.\,\ref{fig:aperture_amplitude}. For example, the core undergoes substrate shifts from AD250 to AD350 then to AD600 with nearly uniform steps of 25\,mm and a corresponding ripple is visible in the aperture fields. If these material junctions contributed to aperture perturbations they would appear to have a period $d$ of roughly $50$\,mm. Periodic aperture perturbations result in farfield sidelobes at $\theta=\sin^{-1}$($\lambda_0$/d) \cite{Brown_1950}, so a $50$\,mm periodic aperture perturbation will produce sidelobes at $8.6$\,$^\circ$ and $11.5$\,$^\circ$ for the $40$\,GHz and $30$\,GHz patterns, respectively. Fig. \ref{fig:gain_patterns} shows the actual sidelobe locations to be at $8$\,$^\circ$ and $10$\,$^\circ$ for these frequencies, reinforcing the claim that the core substrate junctions are perturbing the aperture and producing spurious sidelobes.
Clearly there are other perturbations visible in the aperture (from Fig.\,\ref{fig:aperture_phase} and Fig.\,\ref{fig:aperture_amplitude}) with many appearing only at higher frequencies. This is consistent with the fact that higher frequency operation would be more sensitive to fabrication errors and discontinuities in the realized GRIN profile. The greater aperture nonuniformity that results at higher frequency also explains why the gain reduction from simulation is greater at these frequencies (Fig. \ref{fig:gain_patterns}). 
We considered fabrication nonidealities in subsequent simulations: increased lens thickness, bonding-agent contributions (loss and dielectric interference) and material loss. The maximum gain deviation was less than 0.5\,dB across all bands indicating that other nonidealities which cannot be modeled accurately within simulation (material junctions and permittivity discontinuities, individual perforation tolerances, and individual substrate tolerances on thickness, loss, and dielectric constant) are primarily affecting the lens performance. These nonidealities are entirely related to fabrication and therefore are not expected to appear in the simulated patterns.


\subsection{Wideband Performance and Efficiency} \label{sec:Wideband}

\begin{figure*}
    \centering
    \includegraphics[width=6in]{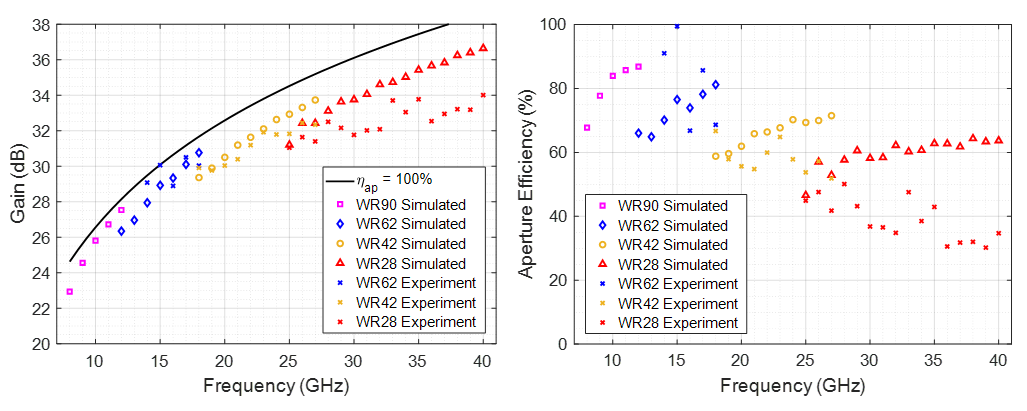}
    \caption{Simulated and experimental maximum gains (left) and corresponding aperture efficiencies (right) in the WR-90, WR-62, WR-42, and WR-28 Bands. The WR-90 and lower half of WR-62 values are simulation only.}
    \label{fig:gains_efficiencies}
\end{figure*}

\begin{figure}
    \centering
    \includegraphics[width=3in]{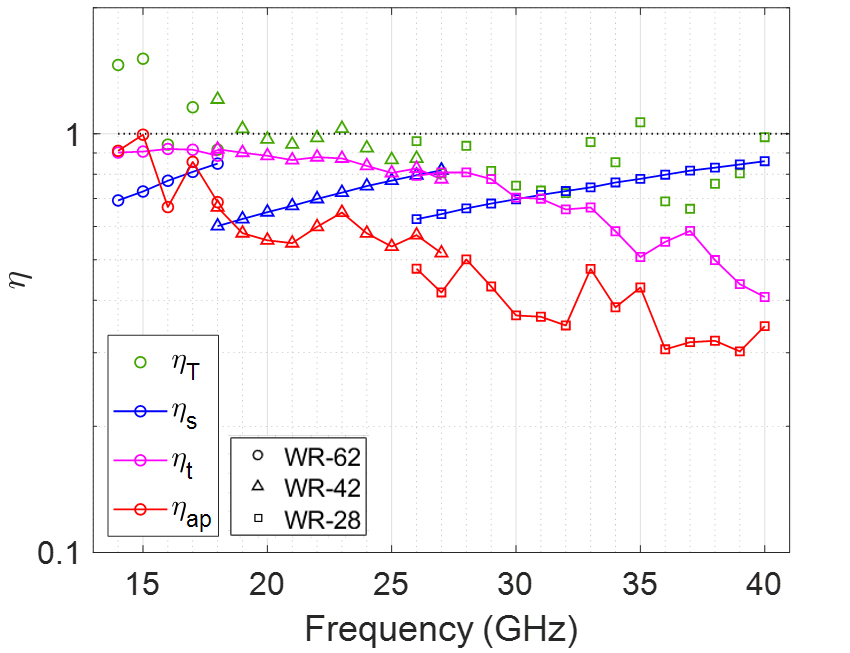}
    \caption{Transmission efficiency ($\eta_{T}$, green trace),\textcolor{black}{\st{illumination efficiency ($\eta_{ill}$, blue trace), phase efficiency ($\eta_{ph}$, magenta trace)} spillover efficiency ($\eta_{s}$, blue trace), taper efficiency ($\eta_{t}$, magenta trace),} and overall aperture efficiency ($\eta_{ap}$, red trace) of the lens across the WR-62 (circle markers), WR-42 (triangle markers), and WR-28 bands (square markers).}
    \label{fig:variousEtas}
\end{figure}

The maximum experimental and simulated gain values and their corresponding aperture efficiencies versus frequency are given in Fig.\,\ref{fig:gains_efficiencies} (left) and (right), respectively. Measurements were performed with WR-62, WR-42, and WR-28 feeds so the measured values are limited to $12.4\,$ to $40\,$GHz. However, simulations were conducted in the WR-90 band (down to $8.2\,$GHz). The gain values generally increase with frequency from around $29$\,dB at $14$\,GHz to $34$\,dB at $40$\,GHz. This gain increase is due to the aperture size becoming electrically larger but it improves at a rate less than $20\log(f)$ indicating that aperture efficiency decreases with frequency. 

It should be noted that the experimental gain values in the WR-62 band are only shown down to $14$\,GHz. This is because the multipath contributions in the time-gated far field chamber obfuscated the measurement below these frequencies and we were not confident in the measurement results. However, simulated results are provided for the remainder of the WR-62 band. We also provide simulated results with a $10$\,dBi WR-90 horn feed covering $8.2$ to $12$\,GHz. Considering that the agreement between measured and simulated gain improves at lower frequencies, we believe these simulated results give a good indication of the lens performance at these frequencies. In addition, these simulations confirm the anticipated breakdown of impedance match at low frequency (see Fig.\,\ref{fig:FresnelTransmission}). Where the other bands show a gradual decline in simulated aperture efficiency with decreasing frequency (due to increased spillover loss), the WR-90 band efficiency falls off steeply below $10$\,GHz which we attribute to the rapid degradation of impedance match in the tapers. While measurement indicates efficient lens operation down to $14\,$GHz, based on these simulations we expect $8$\,GHz to be the lower limit for efficient lens operation.


For a thorough wideband analysis of the lens it is necessary here to decompose the overall aperture efficiency $\eta_{ap}$ into its constituent parts (as outlined by Eq. \ref{eq:ApEff}) and consider each term individually. This is shown graphically in Fig. \ref{fig:variousEtas} where the primary efficiency terms  $\eta_{T}$, \textcolor{black}{\st{$\eta_{ill}$} $\eta_{s}$ and \st{$\eta_{ph}$}$\eta_{t}$} are plotted alongside $\eta_{ap}$ over frequency (note: $\eta_{a}$ and $e_{r}$ in (\ref{eq:ApEff}) are still assumed to be unity and are not included).

\textcolor{black}{\st{Illumination}Spillover} efficiency \textcolor{black}{\st{$\eta_{ill}$} $\eta_{s}$} is shown in the blue trace of Fig. \ref{fig:variousEtas}. These values are calculated based on the power patterns for each feed at each frequency. \textcolor{black}{\st{$\eta_{ill}$} $\eta_{s}$} ranges from $60$\% to \textcolor{black}{\st{$75$\%} $85$\%} across all three bands but is discontinuous between bands due to the variation in gain across a band for any given horn. The horn antennas all have nominally $10$\,dBi of gain but this is only technically true around midband. For example, the $10$\,dBi WR-28 horn exhibits $\approx12.6$\,dB of gain at $40$\,GHz but only $9$\,dB at $26.5$\,GHz. This gain trend follows for all horns and coincides with the trends in \textcolor{black}{\st{$\eta_{ill}$} $\eta_{s}$}. This is because the demonstration lens was optimized for $40$\,GHz and thus the narrowest feed pattern in the WR-28 band. As a result, the lens is biased towards narrower feed patterns and the \textcolor{black}{\st{$\eta_{ill}$} $\eta_{s}$} values increase with frequency in each band. This also explains the discontinuities in \textcolor{black}{\st{$\eta_{ill}$} $\eta_{s}$} in the overlap regions. Here, lower-frequency horns operating at the upper end of their bands out-perform higher-frequency horns operating at the lower ends of their bands. 
\textcolor{black}{\st{Phase efficiency $\eta_{ph}$} Taper efficiency $\eta_{t}$} is calculated based on the aperture fields from the holographic back projections of the measured near-field aperture \textcolor{black}{using the following: 
\begin{equation}
    \eta_t = \frac{1}{A}\frac{|\iint_A E_{ap} ds|^2}{\iint_A |E_{ap}|^2 ds}
\end{equation}
\label{eq:taperEfficiency}}
\noindent \textcolor{black}{where $E_{ap}$ is the complex field distribution in the aperture and $A$ is the aperture area\cite{Stutzman}. $\eta_t$ was also calculated for the fictitious case where the aperture fields have a uniform amplitude (phase) distribution but with phase (amplitude) distribution as measured by the holographic back projection -- this value is denoted $\eta_{t,Ph}$ ($\eta_{t,Mag}$). In this way, the experimental amplitude and phase distributions were treated separately and their effects approximately disambiguated. It was found that for all frequencies $\eta_{t,Mag}$ was near unity and $\eta_t\approx\eta_{t,Ph}$, indicating that $\eta_t$ is dominated by phase non-uniformity in the aperture fields and is then attributed to phase collimation errors and fabrication/material nonidealities.
\st{For this analysis, fictitious aperture fields were created by overlaying the dominant-pol phase information (ie, experimental phase profile) on a uniform amplitude field distribution. This aperture distribution was then propagated to the far field and its maximum gain value determined. In this way, any degradation in gain would be due solely to the phase nonuniformity in the aperture. This degraded gain was then compared to the gain corresponding to $\eta_{ap}$=$1$ and the resulting difference was taken to be $\eta_{ph}$.}} As can be seen in the magenta trace of Fig. \ref{fig:variousEtas}, \textcolor{black}{\st{$\eta_{ph}$} $\eta_t$} ranges from $90$\% to $95$\% in the WR-62 band before decreasing (nearly) monotonically to $45$\% at $40$\,GHz in the WR-28 band. The extremely high efficiencies at lower frequencies indicate that the lens is collimating the feed phase with very little fabrication errors. Considering that the lens is a true time delay component exploiting non-dispersive processes we expect similar performance at higher frequencies. The comparatively low \textcolor{black}{\st{$\eta_{ph}$} $\eta_t$} values at the higher frequencies are then most likely due to material discontinuities and fabrication errors --- nonidealities whose effects are certainly exacerbated by higher frequencies. Broadly speaking, the lens is collimating the feed phase at higher frequencies but the corresponding higher propagation constant compounds phase errors due to fabrication which degrades the \textcolor{black}{\st{aperture phase} taper} efficiency.

The transmission efficiency $\eta_T$ is given in the green traces in Fig. \ref{fig:variousEtas}. These values are calculated by dividing $\eta_{ap}$ by \textcolor{black}{\st{the product of $\eta_{ph}$}} $\eta_{ill}$ -- ie, attributing all remaining gain degradation to the transmission error. These values are consistently high across the whole band of interest indicating that the wideband match functioned much as anticipated. However, these values are lowest in the WR-28 band, typically ranging between $70$\% to $80$\%. Again, this is likely due to fabrication errors and tolerances being exacerbated by frequency, considering that lower frequencies experienced excellent transmission and that the matched taper design is a high-pass response. 

One will also note that in several cases $\eta_T$ is shown to be above unity. This is non-physical and due in large part to the noise in the calculation of $\eta_T$. For example, $\eta_T$ is given as \textcolor{black}{\st{$111$\%} $106$\%} at $35$\,GHz. Assuming all other efficiencies are unity, a gain measurement error of \textcolor{black}{\st{$0.45$\,dB} $0.21$\,dB} would account for the nonphysical ($\>100$\%) transmission efficiency. Between the calculation noise on $\eta_{ill}$ and \textcolor{black}{\st{$\eta_{ph}$} $\eta_t$} and the measurement noise on $\eta_{ap}$, this is not an unreasonable variation. 

The error on $\eta_T$ is more egregious in the lower bands however. The worst example is at $15$\,GHz where \textcolor{black}{\st{$\eta_T=160$\%} $\eta_T=150$\%}. This corresponds to an error of \textcolor{black}{\st{$2$\,dB} $1.8$\,dB} and it's unlikely that this is solely due to calculation/measurement noise. This extremely high $\eta_T$ is related to the extremely high measured $\eta_{ap}$ at this frequency and it is necessary to discuss why the $\eta_{ap}$ values for the WR-62 band are so high.

\begin{table*}
\centering
\ra{1.3}
\begin{threeparttable}
\begin{tabular}{@{}lccccc@{}}
    \toprule
    Ref. & Design & Thickness & Frequency & $\eta_{app}$ & Bandwidth \\
         &        &  (Radii)  &   (GHz)   &     (\%)     & (Ratio)   \\
    \midrule
    \cite{MaCui2013}         & Hemispherical half-Maxwell Fisheye & 1 & 13-18 & 42.6-62.6 & 1.4:1 \\
    \cite{PetosaPerforated}  & Fresnel-zone & 0.08 & 28-32 & 33 & 1.14:1 \\
    \cite{MahmoudKishk_2014} & Fresnel-zone & 0.14 & 90-100 & 20-21 & 1.1:1 \\
    \cite{elef_matched_2017} & Aperture GRIN & 0.04 & 34 & 33 & - \\
    \cite{Budhu2019}         & Shaped GRIN & 0.86 & 13.4 & 43.5 & - \\
    \cite{Imbert2014}        & Flat GRIN & 0.56 & 60,77 & 27,19 & - \\
    \cite{QiCui2013}         & Flat GRIN & 1.04 & 12,15,18 & 59,59,55 & 1.5:1 \\
    \cite{Anastasio2019}     & Flat GRIN & 0.42 & 13.4 & 67.3 & - \\
    {[This work]} & Flat GRIN & 0.3 & \textcolor{black}{(8\tnote{*} )}14-40 & \textcolor{black}{(86\tnote{*} )}72-31 & \textcolor{black}{(5\tnote{*} )}2.9:1 \\
    \bottomrule
    
\end{tabular}
\begin{tablenotes}
\item[*] \textcolor{black}{Simulated with realistic feeds}
\end{tablenotes}
\end{threeparttable}
\vspace{0.1in}
\caption{GRIN lens state of the art}
\label{tab:GRINSOA}
\end{table*}

As discussed earlier, the better low-frequency performance of the lens is in large part due to the reduced sensitivity to fabrication errors. However, this can only contribute so much additional gain --- the low-frequency efficiencies are substantially higher due to swelling of the effective aperture size. That is to say that the close-in spillover radiation near the edge of the lens surface contributes to the total aperture area. At high frequencies, the variation in phase outside the collimated aperture is so rapid that phase uniformity vanishes quickly beyond the physical edge of the lens but at lower frequencies the phase accumulation is sufficiently slow that the phase taper near the physical lens edge retains partial collimation. This actually contributes to a slightly larger radiating aperture. For low frequencies it is difficult to identify a hard boundary for the aperture but using the physical lens surface to calculate aperture efficiency (as is done for Fig.\,\ref{fig:gains_efficiencies} and Fig.\,\ref{fig:variousEtas}) can result in seemingly extreme values (this affect is also clearly seen in the simulated WR-90 results in the magenta square trace of Fig.\,\ref{fig:gains_efficiencies}). This effect has been previously reported in high aperture efficiency GRIN lenses \cite{Anastasio2019}. In our specific case, $\eta_{ap}$ and thus $\eta_T$ are referenced to an 8\," aperture and approach values at or above unity as a result.

\subsection{Comparison with state-of-the-art}

Table\,\ref{tab:GRINSOA} shows the wideband performance of our lens relative to the state-of-the-art. Of the reported lens antennas, the taper lens in this work exhibits both the largest bandwidth ($2.9$:$1$ measured and $5.0$:$1$ simulated) and the highest aperture efficiency bound ($72$\%--$31$\% across the measured band). Aperture efficiency in the table is as reported in the references or computed from available data as, 
\begin{equation}
    \eta_{ap} = G_{meas}/\left(A_p 4\pi/\lambda^2\right),
\end{equation}
\noindent where $G_{meas}$ is the reported measured gain and $A_p$ is the reported physical area of the lens. If a frequency range is given, the efficiency is reported over the same frequency range. In some cases (e.g., \cite{Anastasio2019}) reported frequency range is limited by the feed element and thus a practical lens operating frequency range cannot be fairly compared. The demonstration lens is also among the thinner lenses in the group ($0.3R$), due to the high permittivity contrast materials in its construction. Though far field measurements only indicate gain up to $40$\,GHz it is likely that the lens can operate at higher frequencies as well (with a corresponding reduction in aperture efficiency due to manufacturing errors having a more pronounced effect at higher frequency). Overall Table\,\ref{tab:GRINSOA} indicates that the taper lens antenna presented in this work exhibits high performance across many metrics and the proposed method to simply and rapidly design high-efficiency and wideband GRIN lenses is promising.

\section{Conclusion}

A GRIN lens design methodology utilizing wideband impedance matched unit-cells is proposed based on ray tracing approximations. A demonstration lens is prescribed and fabricated using perforated dielectric structures to build the refractive index profile. The lens is characterized from $14$\,GHz to $40$\,GHz and demonstrates aperture efficiency above $30$\% across all tested frequencies, with aperture efficiencies above $55$\% being demonstrated for frequencies below $25$\,GHz. Specifically, phase and transmission efficiencies for the whole band are above $45$\% and $70$\% respectively, indicating wideband wide-angle transmission and validating the matched unit-cell approach. It should be emphasized that above $33$\,GHz the aperture efficiency is dominated by \textcolor{black}{\st{phase} taper} efficiency but below $33$\,GHz the aperture efficiency is dominated by the \textcolor{black}{\st{illumination} spillover} efficiency which is ultimately the fundamental limit for aperture efficiency. These results suggest that this design method generally is suitable for wideband GRIN lens antennas. That is, given a well-characterized artificial dielectric material library, a matched unit-cell library can be produced that can be used to generate a GRIN lens for any $F/D$, lens diameter, and arbitrary feed pattern---assuming that the library can accumulate sufficient phase over all expected incident angles.  

However, the discrepancies between the simulated and experimental results of the demonstration lens and its associated aperture field distributions indicate that care must be taken to guarantee smooth phase accumulation throughout the body of the lens. Unit-cell arrangement in the lens needs to account for phase perturbations propagating to the aperture as a result of strong material discontinuities. For future lenses, the material junctions in the core could be staggered randomly so that resulting perturbations attenuate quickly and do not perturb the aperture fields. This would eliminate the sidelobes resulting from the periodic aperture disturbances and improve the overall aperture efficiency just by virtue of having smoothed out the amplitude- and phase-distributions in the aperture. In addition, we expect the wideband matching to improve by minimizing similar discontinuities in the tapers. 


In general this design approach is straightforward and the results presented in Fig.\,\ref{fig:variousEtas} and Table\,\ref{tab:GRINSOA} validate its implementation. Process revisions aside, it is our hope that this technique is useful for the design of high aperture efficiency GRIN lenses in the future and at least inspires similar design approaches for wideband impedance matching in GRIN lenses.

\appendices

\section*{Acknowledgment}
The authors would like to thank K. Hersey, Dr. T. Comberiate, and J. McKnight for their commitment to the effort and for the many helpful discussions throughout.


\bibliographystyle{IEEEtran}
\ifCLASSOPTIONcaptionsoff
  \newpage
\fi
\bibliography{TAP_taper_lenses}

\begin{IEEEbiography}
[{\includegraphics[width=1in,height=1.25in,clip,keepaspectratio]{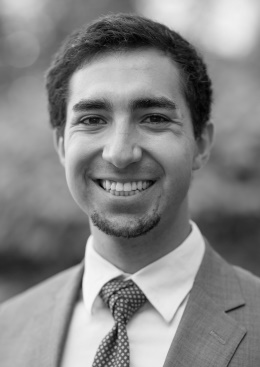}}]
{Nicolas~C.~Garcia} received the B.S. degree in electrical engineering from the University of Notre Dame in 2017. He is currently pursuing a Ph.D. at the same institution. His research interests focus on low-profile gradient index (GRIN) materials and their applications for emerging 5G and millimeter wave antennas and technologies.
\end{IEEEbiography}

\begin{IEEEbiography}[{\includegraphics[width=1in,height=1.25in,clip,keepaspectratio]{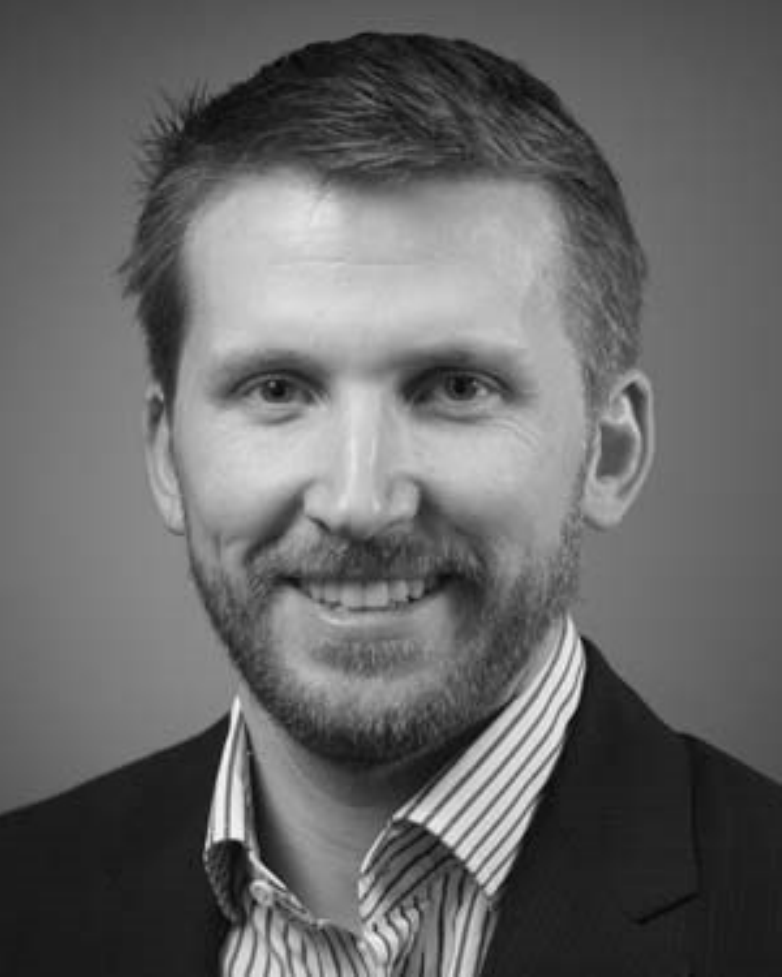}}]{Jonathan~D.~Chisum} (S'02--M'06--SM'17) received the Ph.D. in Electrical Engineering from the University of Colorado at Boulder in Boulder, Colorado USA, in 2011. 

From 2012 to 2015 he was a Member of Technical Staff at the Massachusetts Institute of Technology Lincoln Laboratory in the Wideband Communications and Spectrum Operations groups. His work at Lincoln Laboratory focused on millimeter-wave phased arrays, antennas, and transceiver design for electronic warfare applications. In 2015 he joined the faculty of the University of Notre Dame where he is currently an Assistant Professor of Electrical Engineering. His research interests include millimeter-wave communications and spectrum sensing with an emphasis on low-power and low-cost technologies. His group focuses on gradient index (GRIN) lenses for low-power millimeter-wave beam-steering antennas, nonlinear (1-bit) radio architectures for highly efficient communications and sensing up through millimeter-waves, as well as reconfigurable RF circuits for wideband distributed circuits and antennas. 

Dr. Chisum is a senior member of the IEEE, a member of the American Physical Society, and an elected Member of the U.S. National Committee (USNC) of the International Union or Radio Science's (URSI) Commission D (electronics and photonics). He is the past Secretary and current Vice-chair for USNC URSI Commission D: Electronics and Photonics.

\end{IEEEbiography}

\end{document}